\documentclass[epj]{webofc}
\usepackage[utf8]{inputenc}
\usepackage[varg]{txfonts}   % Web of Conferences font
\usepackage{booktabs}
\usepackage{xcolor}
\definecolor{darkred}{rgb}{0.4,0.0,0.0}
\definecolor{darkgreen}{rgb}{0.0,0.4,0.0}
\definecolor{darkblue}{rgb}{0.0,0.0,0.4}
\usepackage[bookmarks,linktocpage,colorlinks,
    linkcolor = darkred,
    urlcolor  = darkblue,
    citecolor = darkgreen]{hyperref}
\usepackage{subfigure}
\wocname{EPJ Web of Conferences}
\woctitle{Lattice2017}
\def\be{\begin{equation}}
\def\ee{\end{equation}}
\def\bea{\begin{eqnarray}}
\def\eea{\end{eqnarray}}

\definecolor{green}{rgb}{0,.5,0}

\begin{document}
%%%%%%%%%%%%%%%%%%%%%%%%%%%%%%%%%%%%%%%%%%%%%%%%%%%%%%%%%%%%%%%%%%%%%%%%%%%%%
%
\selectlanguage{english}
%----------------------------------------------------------------------------
\title{%
Proton mass decomposition
}
%----------------------------------------------------------------------------
\author{%
\firstname{Yi-Bo} \lastname{Yang}\inst{1}\fnsep\thanks{Speaker, \email{yangyibo@pa.msu.edu}} \and
\firstname{Ying} \lastname{Chen}\inst{2} \and
\firstname{Terrence} \lastname{Draper}\inst{3} \and
\firstname{Jian}  \lastname{Liang}\inst{3} \and
\firstname{Keh-Fei}  \lastname{Liu}\inst{3}
% etc.
}
%----------------------------------------------------------------------------
\institute{%
Department of Physics and Astronomy, Michigan State University, East Lansing, MI 48824, USA
\and Institute of High Energy Physics and Theoretical Physics Center for Science Facilities,\\
Chinese Academy of Sciences, Beijing 100049, China
\and
Department of Physics and Astronomy, University of Kentucky, Lexington, KY 40506, USA
}
%----------------------------------------------------------------------------
\abstract{We report the results on the proton mass decomposition and also on the related quark and glue momentum fractions. The results are based on overlap valence fermions on four ensembles of $N_f = 2+1$ DWF configurations with three lattice spacings and volumes, and several pion masses including the physical pion mass. With 1-loop perturbative calculation and proper normalization of the glue operator, we find that the $u, d,$ and $s$ quark masses contribute 9(2)\% to the proton mass. The quark energy and glue field energy contribute 31(5)\% and 37(5)\% respectively in the $\overline{MS}$ scheme at $\mu = 2$ GeV. The trace anomaly gives the remaining 23(1)\% contribution.
The $u,d,s$ and glue momentum fractions in the $\overline{MS}$ scheme are consistent with the global analysis at $\mu = 2$ GeV.}
%----------------------------------------------------------------------------
\maketitle
%----------------------------------------------------------------------------
\section{Introduction}\label{intro}
The Higgs boson provides the source of quark masses. But how it is related to the proton mass and thus the mass of nuclei and atoms is another question. The masses of the valence quarks in the proton are just $\sim$3 MeV per quark which is directly related to the Higgs boson, while the total proton mass is 938 MeV. How large the quark and gluon contributions to the proton mass are, is a question that can only answered by solving QCD nonperturbatively, and/or with information from experiment. With phenomenological input, the first decomposition was carried out by Ji over twenty years ago~\cite {Ji:1994av}. As in Ref.~\cite{Ji:1994av,Yang:2014xsa}, the Hamiltonian of QCD can be decomposed as
 %
 % [I have simplified the following equations for the proceedings. We can explained it in more detail in a paper.]
 \be 
 M = - \langle T_{44} \rangle =\langle H_m\rangle + \langle H_E\rangle (\mu)+ \langle H_g\rangle (\mu) +   \langle H_a \rangle, \label{eq:total}
\ee
and the trace anomaly gives
\be
M = - 4\,\langle \hat{T}_{44} \rangle= \langle H_m\rangle +  4\,\langle H_a\rangle,\label{eq:anomly}
\ee
with $H_m$, $H_E$, and $H_g$ denoting the contributions from the quark mass, the quark energy, and  the glue field energy, respectively:
\bea
  H_m &=& \sum_{u,d,s\cdots}\int d^3x\, m\, \overline \psi  \psi,   \quad\
  H_E = \sum_{u,d,s...}\int d^3x~\overline \psi(\vec{D}\cdot \vec{\gamma})\psi,\quad  \nonumber\\
%H_q = H_E+H_m&, \quad
 H_g &=& \int d^3 x~ {\frac{1}{2}}(B^2- E^2), 
 \eea
and the QCD anomaly term $H_a$ is the joint contribution from the quantum anomaly of both glue and quark,
 \bea
 H_a= H_g^a +H^{\gamma}_m, \quad
 H_g^a =\int d^3x~ \frac{-\beta(g)}{4g}( E^2+ B^2), \quad
H^{\gamma}_m=\sum_{u,d,s\cdots}\int d^3x\, \frac{1}{4}\gamma_m m\, \overline \psi  \psi.
 \eea
All the $\langle H \rangle$ are defined by $\langle N|H |N\rangle/\langle N|N\rangle$ where $|N\rangle$ is the nucleon state in the rest frame.
Note that $\langle H_E+H_g\rangle$, $\langle H_m\rangle$ and $\langle H_a\rangle$ are scale and renormalization scheme independent, but 
$\langle H_E\rangle (\mu)$ and $\langle H_g\rangle (\mu)$ separately have scale dependence. 

%The above decomposition requires the calculation of the energy momentum tensor matrix elements in the proton, and the attempts to construct %a conserved stress tensor on the lattice have been made perturbatively and non-perturbatively \cite{Caracciolo:1989pt} and recently by Suzuki %\cite{Suzuki:2013gza,Makino:2014taa} with Wilson flow at finite lattice spacing. Recently, the first calculation of the S-wave meson mass %decomposition has been carried out with lattice QCD, while just the quark components are calculated directly and the glue ones are deduced %from the quark parts~\cite{Yang:2014xsa}.

The nucleon mass $M$ can be calculated from the nucleon two-point function. If one calculates
further $\langle H_m\rangle$ and $\langle H_E\rangle (\mu)$, then $\langle H_g\rangle (\mu)$ and $\langle H_a\rangle$ can be obtained through Eqs.~(\ref{eq:total}) and (\ref{eq:anomly}). The approach has been adopted to decompose the S-wave meson masses to gain insight about
contributions of each term from light mesons to charmonuims~\cite{Yang:2014xsa}. In this work, the quark energy $\langle H_E\rangle$
is obtained from the quark momentum fraction from a local current and $\langle H_m\rangle$ and with the help of the equation of motion, i.e. 
\be    \label{eq:H_E}
\langle H_E \rangle =\frac{3}{4} \langle x \rangle_q M-\frac{3}{4}\langle H_m\rangle.
\ee
Since there is an $\mathcal{O}(a^2)$ error in the equation of motion due to the fact that the local energy-momentum tensor operator adopted is
not the conserved current, the concomitant systematic error can be up to 20\% for the light quark case in the meson mass study of the 
pseudoscalar meson~\cite{Yang:2014xsa}. In principle, it would be better to use the conserved energy-momentum tensor (EMT) on the lattice
to avoid the need for normalization and attempts to construct such a conserved EMT on the lattice have been made perturbatively and non-perturbatively~\cite{Caracciolo:1989pt} and recently by Suzuki \cite{Suzuki:2013gza,Makino:2014taa} with gradient flow at finite lattice spacing. However, they are complicated to construct. In the present work, we still use the local current and will address the normalization
in addition to renormalization and mixing.

%but there are up to 20\% systematic errors for the light quark case due to the lattice regularization effect which breaks the quark's equation of %motion, at finite lattice spacing. This breaking gives $\langle H_E\rangle$ an additional mixing with $\langle H_m\rangle$ (and then also that %with $H_a$ in the 2-loop level) under the lattice regularization. At the same time, $\langle H_E\rangle$ is also mixed with $\langle %H_g\rangle$. So the renormalization of these four components, $\langle H_E\rangle$, $\langle H_m\rangle$, $\langle H_g\rangle$ and $\langle %H_a\rangle$, is a $4\times4$ matrix which is highly non-trivial to calculate under the lattice regularization.
%However, there is a way to avoid the direct calculation of this 4-by-4 mixing.

%Since the quark mass term is less than 10\% of the proton mass~\cite{Durr:2015dna,Yang:2015uis,Abdel-Rehim:2016won,Bali:2016lvx} and 
%$\gamma_m\ll 1$, we can access $\langle H^a_g\rangle$ by $\langle H_a\rangle$ with Eq.~\ref{eq:anomly} with a tiny correction and 
%thus avoid the difficulty in its renormalization. 

In addition to calculating the quark momentum  fraction $\langle x\rangle_q$, we also calculate the glue momentum fraction
$\langle x \rangle_g$. The latter is directly related to the glue field energy
\be  \label{eq:H_g}
\langle H_g \rangle =\frac{3}{4} \langle x \rangle_g M.
\ee
We will discuss the normalization, renormalization, and mixing of $\langle x\rangle_q$ and $\langle x \rangle_g$ later. 

In this proceeding, we will calculate the renormalized quark and glue momentum fractions in the proton on four lattice ensembles and interpolate the results to the physical pion mass with a global fit including finite lattice spacing and volume corrections. Then we will combine the previous $\langle H_m\rangle$ result~\cite{Yang:2015uis} to obtain the full decomposition of the proton mass. %The bare value of $\langle H^a_g\rangle$ will be also presented to compare with the value deduced by Eq.~\ref{eq:anomly}.

%----------------------------------------------------------------------------
\section{Numerical setup}\label{sec-1}

We use overlap valence fermion on $(2+1)$ flavor RBC/UKQCD DWF gauge configurations from four ensembles on $24^3\times64$ (24I),  $32^3\times64$ (32I)~\cite{Aoki:2010dy}, $32^3\times64$ (32ID) and $48^3\times96$ (48I)~\cite{Blum:2014tka} lattices.  These ensembles cover three values of the lattice spacing and volume, and four values of the quark mass in the sea, and then allow us to implement a global fit on our results to control the systematic uncertainties as in Ref.~\cite{Yang:2015uis,Sufian:2016pex}. Other parameters of the ensembles used are listed in Table~\ref{table:r0}. 

\begin{table}[htbp]
\begin{center}
\caption{\label{table:r0} The parameters for the RBC/UKQCD configurations\cite{Blum:2014tka}: spatial/temporal size, lattice spacing, the sea strange quark mass under $\overline{MS}$ scheme at {2 GeV}, the pion mass with the degenerate light sea quark (both in unit of MeV), and the number of configurations used in this proceeding.}
\begin{tabular}{ccccccc}
Symbol & $L^3\times T$  &a (fm)  &$m_s^{(s)}$&  {$m_{\pi}$}   & $N_{cfg} $ \\
\hline
24I & $24^3\times 64$& 0.1105(3) &120   &330  & 203  \\
32I &$32^3\times 64$& 0.0828(3) & 110   &300 & 309  \\
32ID &$32^3\times 64$& 0.1431(7) & 89.4   &171 & 200  \\
48I &$48^3\times 96$& 0.1141(2) & 94.9   &139 & 81  \\
\hline
\end{tabular}
\end{center}
\end{table}

The effective quark propagator of the massive
overlap fermion is the inverse of the operator $(D_c + m)$~\cite{Chiu:1998eu,Liu:2002qu}, where  $D_c$ is chiral, i.e. $\{D_c, \gamma_5\} = 0$ \cite{Chiu:1998gp} and its detailed definition can be found in our previous works~\cite{Li:2010pw,Gong:2013vja,Yang:2015zja}. We used 5 quark {masses} from the range $m_{\pi}\in$(250, 400) MeV on the 24I and 32I ensembles, and 6 quark 
masses from $m_{\pi}\in$(140, 400) MeV on the other two ensembles which have larger volumes and thus allow a lighter pion mass with 
the constraint $m_{\pi}L>3.8$.  For all the quark propagators, 1 step of HYP smearing is applied on all the configurations to improve the signal. Numerical details regarding the calculation of the overlap operator, eigenmode deflation in inversion of the quark matrix, and the $Z(3)$ grid smeared source with low-mode substitution (LMS) to increase statistics are given in~\cite{Li:2010pw,Gong:2013vja,Yang:2015zja}.

The matrix elements we need are obtained from the ratio of the three-point function to the two-point function
\begin{eqnarray}   \label{eq:ratio}
R(t_f,t)&=&\frac{\langle 0|\int d^3 y\Gamma^e{\chi}(\vec{y},t_f){\cal O}(t)\sum_{\vec{x}\in G}\bar{\chi}_S(\vec{x},0)|0 \rangle}{\langle 0|\int d^3 y\Gamma^e\chi(\vec{y},t_f)\sum_{\vec{x}\in G}\bar{\chi}_S(\vec{x},0)|0 \rangle},
\end{eqnarray}
where $\chi$ is the standard proton interpolation field and $\bar{\chi}_S$ is the field with gaussian smearing applied to all three quarks. All the correlation functions from the source points $\vec{x}$ in the grid $G$ are combined to improve the the signal-to-noise ratio (SNR).  ${\cal O}(t)$ is the current operator located at time slice $t$  and $\Gamma^e$ is the unpolarized projection operator of the nucleon. When $t_f$ is large enough, 
 $R(t_f,t)$ approaches the bare nucleon matrix element matrix element $\langle N|{\cal O}|N\rangle$.

For each quark mass on each ensemble, we construct $R(t_f,t)$
 for several sink-source separations $t_f$ from
0.7 fm to 1.5 fm and all the current insertion times
$t$ between the source and sink, combine all the data to do the two-state fit, and then obtain the matrix elements we want with the excited-states contamination under control. The more detailed discussion of the simulation setup and the two-state fit can be found in our previous work~\cite{Yang:2015uis,Sufian:2016pex,Yang:2016plb}.

To improve the signal in the disconnected insertion case needed for the gluon, strange and also the light sea quarks cases, all the time slices are looped over for the proton two-point functions. With 5 steps of the HYP smearing, the signal of the glue operator is further improved as evidenced in Ref.~\cite{Yang:2016plb}. \\\\

\section{Results}\label{sec-2}

The quark and gluon momentum fractions in the nucleon can be defined by the traceless diagonal part of the EMT matrix element in the rest frame~\cite{Horsley:2012pz},
\bea
\langle x \rangle^{\textrm{tr}}_{q,g}&\equiv&\frac{\textrm{Tr}[\Gamma^e\langle N |\frac{4}{3}\bar{T}^{q,g}_{44} |N \rangle]}{M\textrm{Tr}[\Gamma^e\langle N|N \rangle]},\\
\bar{T}^{q}_{44}&=&\int d^3x \overline{\psi}(x) \frac{1}{2}(\frac{3}{4}\gamma_4\overleftrightarrow{D}_4 -\frac{1}{4}{\displaystyle\sum_{i=1,2,3}} \gamma_i\overleftrightarrow{D}_i) \psi (x), \ \ 
\bar{T}^{g}_{44}=\int d^3x \frac{1}{2}(B(x)^2-E(x)^2),\nonumber
\eea
where $M$ is the proton mass, or alternatively by the off-diagonal part of the EMT matrix elements,
\bea
\langle x \rangle^{\textrm{off}}_{q,g}&\equiv&\frac{\textrm{Tr}[\Gamma^e\langle P |T^{q,g}_{4i} |P \rangle]}{P_i\textrm{Tr}[\Gamma^e\langle P|P \rangle]}\\
T^{q}_{4i}&=&\int d^3x \overline{\psi}(x) \frac{1}{4}\gamma_{\{4}\overleftrightarrow{D}_{i\}}\psi (x),\ \ 
T^{g}_{4i}=\int d^3x \epsilon_{ijk} E_j(x)B_k(x),\nonumber
\eea
where $|P\rangle$ is the nucleon state with momentum $P$ and $P_i$ is a non-zero component of $P$. These two definitions should give the same result in the continuum due to the rotational symmetry. But they can be different under the lattice regularization which breaks this symmetry and should be renormalized separately to get consistent results.

In Ref.~\cite{Yang:2016xsb}, we provided the 1-loop renormalization and mixing calculation of $\bar{T}_{44}$ and $\bar{T}_{4i}$. The rotational symmetry breaking effects in the renormalization constant of the quark operator and the mixing from quark to gluon are small, while that in the gluon to quark mixing case is large.
%enough to cause the sign of the mixing coefficient in the $\bar{T}_{44}$ case to be different from that in the $\bar{T}_{4i}$ case. 
The glue renormalization constant turns out to be $\sim$2 at the 1-loop level and is thus not reliable. The renormalization condition provided in Ref.~\cite{Yang:2016xsb} can also be used for the non-perturbative renormalization calculation, and the preliminary result shows that the renormalization constant of the gluon operator with 1-step HYP smearing is about 1.3~\cite{yang:2017glue}. That with more steps of the HYP smearing is under investigation and would be closer to 1, since the corresponding bare gluon matrix elements are slightly increased compared to the 1-step HYP smearing case.

 \begin{figure}[h]
\begin{center}
    \includegraphics[scale=0.3]{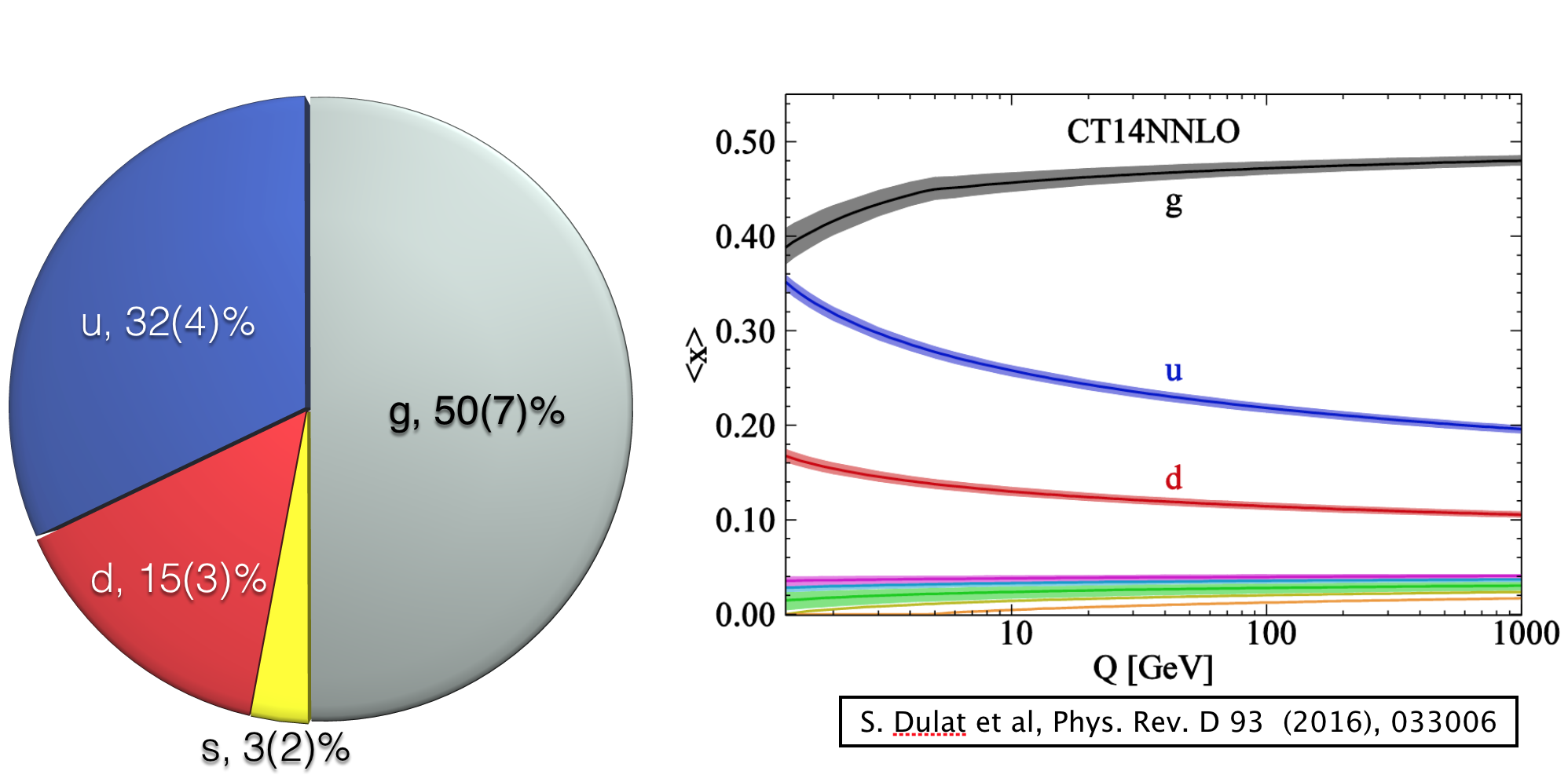}
\end{center}
\caption{The contributions of different quark flavors and glue to the proton momentum fraction. The left panel shows the lattice results renormalized in the  $\overline{\textrm{MS}}$ scheme at 2 GeV with 1-loop perturbative calculation and proper normalization of the glue.
The experimental values are illustrated in the right panel, as a function of the $\overline{\textrm{MS}}$ scale. Our results agree with the experimental values at 2 GeV.}
\label{fig:1}
\end{figure}

In view of the uncertainty in the glue renormalization, we calculate the renormalized quark momentum fractions with the 1-loop perturbative calculation including the mixing of the bare glue momentum fraction and apply the momentum sum rule to determine the renormalized glue momentum fraction. The resulting renormalized momentum fractions of the $u$, $d$, $s$ quarks, and glue in the $\overline{\textrm{MS}}$ scheme at 2 GeV are illustrated in the left panel of Fig.~\ref{fig:1}, while the right panel shows the corresponding experimental values as a function of $Q$~\cite{Dulat:2015mca}. We note that they agree with each other well within uncertainties.

\begin{figure}[h]
\begin{center}
    \includegraphics[scale=0.4]{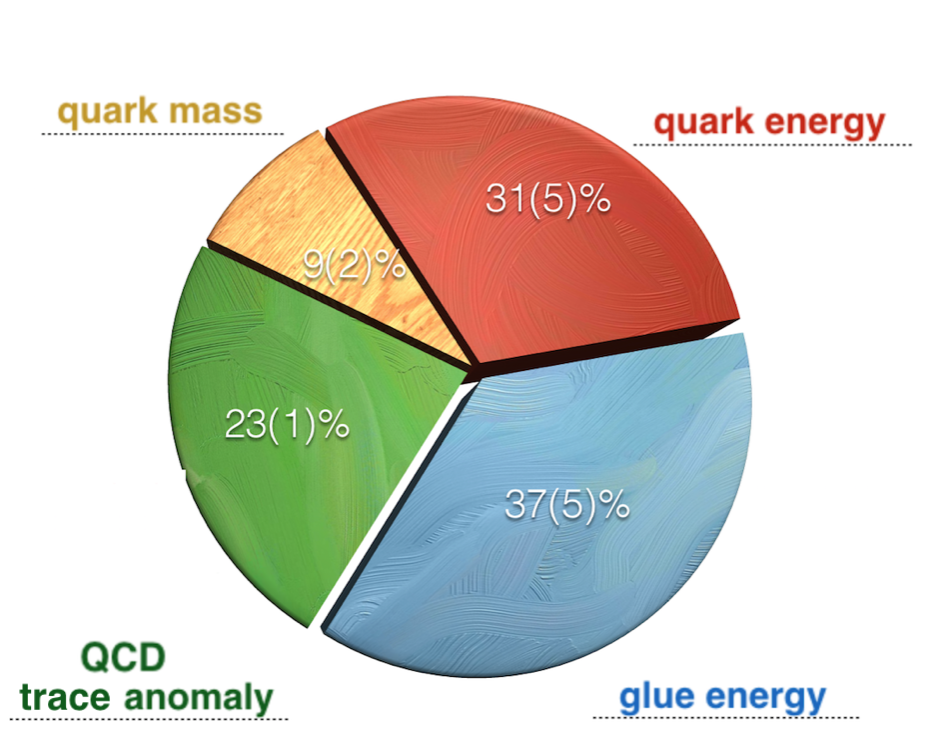}
\end{center}
\caption{The pie chart of the proton mass decomposition, in terms of the quark mass, quark energy, glue field energy and trace anomaly.}
\label{fig:2}
\end{figure}

With these momentum fractions, we can apply Eqs.~(\ref{eq:H_E}) and (\ref{eq:H_g}) to obtain the quark and glue energy contributions in the proton mass, and combine with the quark mass contribution ~\cite{Yang:2015uis} to obtain the entire picture of the proton mass decomposition, as illustrated in Fig.~\ref{fig:2}.\\\\

%----------------------------------------------------------------------------
\section{Summary}\label{sec:details}
In summary, we present a simulation strategy to calculate the proton mass decomposition. The renormalization and mixing between the quark 
and glue energy can be calculated perturbatively or non-perturbatively, while the quark mass contribution and the trace anomaly are renormalization group invariant.  Based on this strategy, the lattice simulation is processed on four ensembles with three lattice spacings and volumes, and several pion masses including the physical pion mass, to control the systematic uncertainties. With 1-loop perturbative calculation and proper normalization on the glue, we obtained the proton mass decomposition, with the quark mass and trace anomaly contributing
9(2)\% and 23(1)\% respectively, while the fractional contributions of the quark and glue field energies are 31(5)\% and 37(5)\% in the
$\overline{\textrm{MS}}$ scheme at 2 GeV. As a check of validity of the present calculation, we find that
the individual $u,d,s$ and glue momentum fractions compare favorably with those from the global fit in the
$\overline{\textrm{MS}}$ scheme at 2 GeV.

\bibliography{reference.bib}

%%%%%%%%%%%%%%%%%%%%%%%%%%%%%%%%%%%%%%%%%%%%%%%%%%%%%%%%%%%%%%%%%%%%%%%%%%%%%
\end{document}